\documentclass[12pt,preprint]{aastex}

\slugcomment{To appear in ApJ, December, 2004}

\shorttitle{Stellar Temperature Effects on YSO Colors}
\shortauthors{B. Whitney et al.}

\begin{document}

\title{2-D Models of Protostars:  III.  Effects of Stellar Temperature}


\author{Barbara. A. Whitney,\altaffilmark{1} 
R\'{e}my Indebetouw,\altaffilmark{2}
J. E. Bjorkman,\altaffilmark{3}
\&
Kenneth Wood,\altaffilmark{4}
}

\altaffiltext{1}{Space Science Institute, 4750 Walnut St. Suite 205,
Boulder, CO 80301; bwhitney@spacescience.org}

\altaffiltext{2}{University of Wisconsin-Madison, Dept. of Astronomy,
475 N. Charter St., Madison, WI 53706; remy@astro.wisc.edu}

\altaffiltext{3}{Ritter Observatory, MS 113, Department of Physics \&
Astronomy, University of Toledo, Toledo, OH 43606-3390;
jon@physics.utoledo.edu}

\altaffiltext{4}{School of Physics \& Astronomy, University of St Andrews, 
North Haugh, St Andrews, Fife, KY16 9AD, Scotland the Brave; 
kw25@st-andrews.ac.uk}

\begin{abstract}

We model how the mid-infrared colors of Young Stellar Objects (YSOs)
vary with stellar temperature.  
The spectral energy distribution (SED) of each object has
contributions from thermal emission of circumstellar dust, from direct
stellar photospheric emission, and from scattered stellar emission.
We first isolate the effects of stellar contributions (direct+scattered) to the SED
using homologous ``Class I'' models: the distribution of circumstellar matter is
chosen to scale with stellar temperature $T_\star$ such that the shape
of the thermal contribution to the SED remains constant.  The relative
contribution of stellar direct and scattered light varies with
$T_\star$, changing the $1-10 \mu$m (mid-infrared; MIR) colors.
Stellar light contributes more to the MIR emission of YSOs with lower
temperature stars ($T_\star \sim$~4000~K) because the emission peak wavelength of
the star is closer to that of the thermal radiation.  In
YSOs with hotter central stars, since the peak of the stellar and thermal spectra
are more separated in wavelength, the $1-10 \mu$m spectrum is closer
to a pure thermal spectrum and the objects are redder.

Next we consider realistic Class 0, I, and II source models and find
that the other dominant effect of varying stellar temperature on YSO SEDs is that of
the inner disk wall:  In high-$T_\star$ models, the dust destruction radius is much
further out with a consequently larger inner disk wall that contributes
relatively more to the $2-10\mu$m flux.
This effect partially offsets that of the stellar contribution leading to
varying behaviors of the $2-10\mu$m flux:  In Class 0 sources, the
trend is for higher $T_\star$ models to have redder colors. 
In Class I sources, the trend applies with some exceptions.
In Class II sources, $2-10\mu$m colors
become redder going from $T_\star=4000$ to 8000 due to decreasing stellar
contribution at $T_\star = 8000 K$, and then become blue again from
8000 to 31500 K due to increasing inner disk wall contribution.  Near edge-on inclinations, the
color behavior is completely different.  

Our modeled MIR protostellar colors have implications for
interpretations of {\it Spitzer } IRAC observations of star formation
regions: It is commonly assumed that the slope of the SED at $1-10\mu$m
is directly related to evolutionary state.  We show that inclination effects,
aperture size, scattered light, and stellar temperature cause a broad spread
in the colors of a source at a single evolutionary state.
Color-magnitude diagrams can
help sort out these effects by separating sources with different $T_\star$
based on their different brightness (for sources at the same distance).

\end{abstract}


\section{Introduction}

With the launch and successful operation of the {\it Spitzer Space
Telescope} \citep{werner04}, there is a wealth of mid-infrared data being collected on
star formation regions near and far
\citep[e.g.][]{allen04,megeath04,reach04,whitney04}. 
The IRAC camera
\citep{fazio04} provides unprecedented sensitivity and mapping speed
using four filters centered on 3.6, 4.5, 5.8, and 8 $\mu$m, and the
MIPS camera \citep{rieke04} is functioning particularly well in the
24$\mu$m band \citep[e.g.][]{muzerolle04}.  Furthermore, many current
ground-based facilities and recent space based facilities (e.g. ISO)
are optimized to collect data in the near/mid-IR.  Thus, in many
cases, scientific analysis of star forming clusters is based only on
broad-band colors in the 1-30$\mu$m range.

Traditionally the slope of the SED in this region, parameterized by
the spectral index $\alpha$ ($=d \log \lambda F_\lambda / d \log \lambda$)
is used to classify evolutionary state \citep{lada87,lada99}.  A
sequence of evolutionary classes is fairly well-defined for protostars
of moderate mass ($M<1-2M_\sun$).  Sources for which $\alpha$ is positive
are thought to have large
(thousands of AU) infalling envelopes and are classified as Class 0 to
Class I. Class 0 sources are very young ($10^4$ yrs) protostars (Andr\'e et al. 1993)
with high infall rates and very collimated outflows.  Class I sources
are older ($10^5$ yrs) with lower infall rates and larger bipolar
cavities carved by meandering jets and molecular outflows 
(G\'omez et al. 1997; Richer et al. 2000; Reipurth et al. 2000).  
Sources with $\alpha<0$ are identified as Class II sources,
pre-main sequence stars surrounded by flared accretion disks.
Obviously there are intermediate stages of evolution between Class I
and II and it is commonly thought that the spectral index continuum is
also a measure of the evolutionary continuum between Class 0 and II
\citep{kenyon95}.

Contrasting with the well-defined class system for low-mass stars, the
evolution of high-mass protostars is poorly understood.  
Disk accretion is more
difficult to model theoretically in the case of a high-mass protostar
because of strong radiation pressure from the central source, high
accretion rates, and large disk masses leading to instabilities.
Some theorists have
suggested protostellar coalescence as an alternative formation
mechanism \citep{bonnell}, but others have succeeded in producing
viable accretion models \citep{behrend,maeder}.  The large volume of
mid-infrared data currently being collected in protostars of all
masses makes it important to understand the infrared properties
of high-mass protostars.  In particular, does the slope of the SED
relate to evolutionary state, as for low-mass stars, and what other
effects are present that should be taken into account when analyzing
the data?

This paper is part of a series on using 2-D radiative transfer models
to interpret data of Young Stellar Objects (YSOs).  In previous papers
(Whitney et al. 2003a,b; Papers I \& II), we showed that inclination
effects can blur the separation of evolutionary states in mid-IR
color-color diagrams of low-mass YSOs.  As an example, in edge-on
Class I sources, the mid-IR flux is dominated by scattered light due
to the large extinction in the disk blocking all stellar and inner
disk radiation; this leads to mid-IR colors that are bluer than
Class II sources.  In contrast, pole-on Class I sources are less red than
average due
to lower extinctions in their partially-evacuated bipolar cavities.
Color-magnitude diagrams and near-IR polarization measurements can
help sort out this blending since both flux and polarization vary with
inclination (e.g., edge-on Class I and II sources will be blue, faint
and have high near-IR polarization).

In this paper, we investigate the variation in mid-IR colors of
YSOs due to the temperature ($T_\star$), or mass, of
the central star.  We assume that disk accretion occurs for stars of
any mass, with physical properties that scale with mass and agree with
observations.  
We find that varying the stellar temperature has two competing effects
on the MIR colors of YSOs.
The first is the relative contribution of stellar scattered+direct flux
to the 1-10 $\mu$m flux for sources with differing $T_\star$.
We demonstrate this effect by showing a
set of homologous ``Class I'' models in \S2.2 in which the thermal spectral shapes
are very similar between models with four different stellar
temperatures.  The disk geometries in these high-$T_\star$ models are unrealistic
because they are chosen to scale homologously with the low-$T_\star$ disk.
\S2.3 shows Class I models with more realistic disk properties in the high-$T_\star$
models.  These models illustrate a second
effect on the MIR SEDs, that of the 
increasing inner disk wall contribution from the higher $T_\star$
sources.  This partially offsets the reddening effect of the stellar
contribution in Class I sources.  \S2.4 extends the models to Class 0 and II sources
and shows that the inner disk wall effect is even more important in Class II
sources, not surprisingly.  
We show color-color diagrams in \S2.5 which show overlap between evolutionary
states due to inclination and stellar temperature; however the color-magnitude
diagrams provide some guidance in separating effects of stellar temperature.
A brief summary is presented in \S3.


\section{Models}


\subsection{Radiative Transfer}

\begin{figure}
\figurenum{1}
\epsscale{1.0}
\plotone{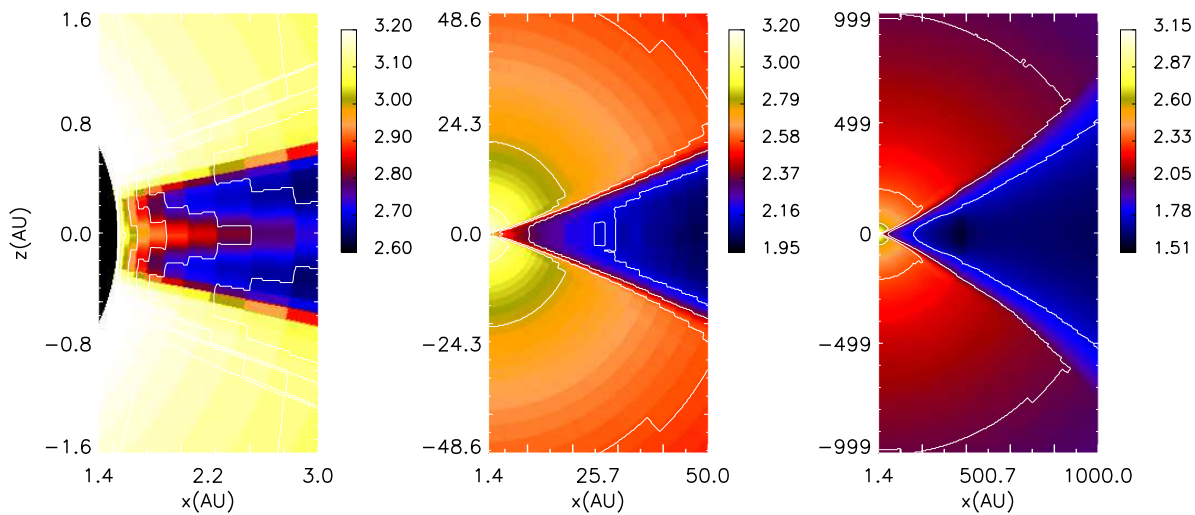}
\caption{\label{temp} 2-D temperature structure of a Class I source (Table 2, with $T_\star$
equal to 8000 K) on three different
size scales.  The small-scale plot (left) shows the hot cavity, the hot inner disk wall and surface, and the effect of accretion heating up the midplane at small radii.
The black region at left is the region inside the dust destruction
radius.  
On larger scales (right panel), we can see 
the shadowing in the equatorial plane caused by the opaque disk.
The color bars show log of the temperature.
The contours correspond to the labeled values on the color bars.
}
\end{figure}

We use a 3-D Monte Carlo Radiative transfer code\footnote{Source code, 
instructions for running, and
sample plotting tools are available at
http://gemelli.spacescience.org/$\sim$bwhitney/codes/codes.html}
described in Paper I, which uses the radiative equilibrium method developed by Bjorkman
\& Wood (2001).  The
geometries considered in this paper are 2-D so we use a 2-D grid
($r-\theta$).  The model geometries consist of a stellar source, a
flared accretion disk, and a rotationally-flattened infalling envelope
(Ulrich 1976; Terebey, Shu \& Cassen 1984) with partially-evacuated
bipolar cavities (Papers I \& II).
The envelope density decreases via a power-law at large radii ($\rho \propto r^{-1.5}$)
and then merges with the ambient density of the surrounding molecular cloud.
Luminosity is generated by the central star and
accretion in the disk.  This radiation is scattered and reprocessed by
the surrounding circumstellar dust.
We note that we solve for the 3-D temperature in any specified
circumstellar geometry; therefore we naturally compute a hot surface
on the inner disk wall as shown in Figure \ref{temp}, as we have in
all our previous publications using these models (e.g., Wood et al. 2002a,b,
Papers I \& II, Rice et al. 2003, Walker et al. 2004).  In addition, our cavity
dust is hot and has a high emissivity, despite its low density (Paper
I, Figure 7).  Therefore our Class 0-II models include
mid-IR thermal emission from the warm disk and cavity regions, in addition
to any envelope component.  Our
models also accurately compute scattering and polarization using
arbitrary scattering phase functions.  Our models conserve flux
absolutely (that is, to 0\%).  The primary source of error in our models
is photon counting statistics (and, when comparing to data, knowledge
of the appropriate input circumstellar geometry and dust properties).
Running more photons produces higher signal-to-noise spectra.  The
models produced for this paper took 3 hours each to run (on 2 GHz PCs
running g77).  The exiting photons were binned into 10 inclinations and 200
frequencies to produce SEDs.

\subsubsection{Dust Properties and Dust Sublimation Radius}

We use similar grain properties as in Paper II (Table 3): a
large-grain model for the high-density regions in the disk (Wood et
al. 2002b); a medium-sized grain model for the upper
layers of the disk (Cotera et al. 2001); and for the envelope and
cavity, a grain model that gives an extinction curve typical of
molecular clouds with $r_V$, the ratio of total-to-selective
extinction, equal to 4.3.  The dust sublimation temperature is chosen
to be 1600~K.

The disk dust sublimation radius was calculated through iteration by running the 
code several times and setting the opacity to be zero in grid cells when the temperature
rises above $T_{sub}$.  In the final iteration, there are no cells with $T > T_{sub}$
(Walker et al. 2004).   
We determined an empirical 
formula that fit our range of stellar temperatures:
\begin{equation}
R_{sub}/R_\star=(T_{sub}/T_\star)^{-2.085}.
\end{equation}
This is very similar to the optically {\it thick} blackbody radiative equilibrium
temperature, $r \propto T^{-2}$, and thus likely has little dependence on the dust opacity
law, unlike the optically thin radiative equilibrium limit \citep{lamers99,beckwith90}.
This behavior is obviously due to the fact that the inner disk wall is opaque at all
wavelengths; this is true over a wide range of disk masses (Wood et al. 2002b).

\subsection{Homologous Models}

We start by specifying a Class I model for a low-mass (0.5 M$_\sun$) YSO, based on
previous observations and models (Kenyon et al. 1993a,b,
Whitney et al. 1997, Lucas \& Roche 1997, 1998; Padgett et al. 1999).  Then we
will scale this model homologously for different stellar temperatures.
To construct a homologous model, we require that the inner and outer
radii be scaled to the dust destruction radius and that the optical
depths (at a given inclination angle) be the same for all the models
(Ivezi\'c \& Elitzur 1997; Carciofi et al. 2004).  This will result in a
homologous temperature distribution for the circumstellar dust, and
the shape of the thermal contribution to the SED will be invariant
(only scaled by the increased luminosity).  Table 1 shows the model
parameters that result.  For all the models we choose the inner radius to
be the dust destruction radius.   For the low-mass model, the outer
radius is 3000 AU (chosen to be as small as reasonable since the high
temperature model will be huge).  The disk radius and envelope
centrifugal radius is 300 AU, and the envelope infall rate is
$6.7\times 10^{-6}$ M$_\sun$/yr.  The model includes a flared disk of mass 0.01
$M_\sun$.  The ratio of disk scale height to radius at the disk outer
radius is $h/r = 0.12$ (chosen to match the HH30 disk, Burrows et
al. 1996; Wood et al. 1998).  For simplicity in comparing models, we
set the disk accretion rate to 0, so the disk radiates from thermal
reprocessing of starlight only.  The bipolar cavity has a curved shape ($z = a (x^2+y^2)^b$,
where $b=1.5$ and $a$ is set by the opening angle)
and an opening angle of 20 degrees at the outer radius.  The
bipolar cavity is filled with constant-density dust with an optical
depth along the polar direction of $\tau_V=5$ at V (0.55 $\mu$m).
Because the low-$T_\star$ models have smaller outer envelope radii,
they have correspondingly higher density in the bipolar cavities.
Based on these parameter choices, the optical depth through the
envelope at an inclination of 60 degrees is $\tau_V$=25; and through
the disk midplane is $\tau_V$=193,000.

The parameters for the other models are then chosen to give the same
inner and outer radii scaled to dust destruction radius, and the same
optical depths at 0\arcdeg (cavity), 60\arcdeg (envelope), and 90\arcdeg (disk).
Thus the envelope and disk masses and radii grow with the higher
$T_\star$ models, as shown in Table 1.  We choose stellar parameters
appropriate for young stars of age $3 \times 10^5$ yrs \citep{siess}.

\begin{figure}
\figurenum{2a}
\epsscale{1.0}
\plotone{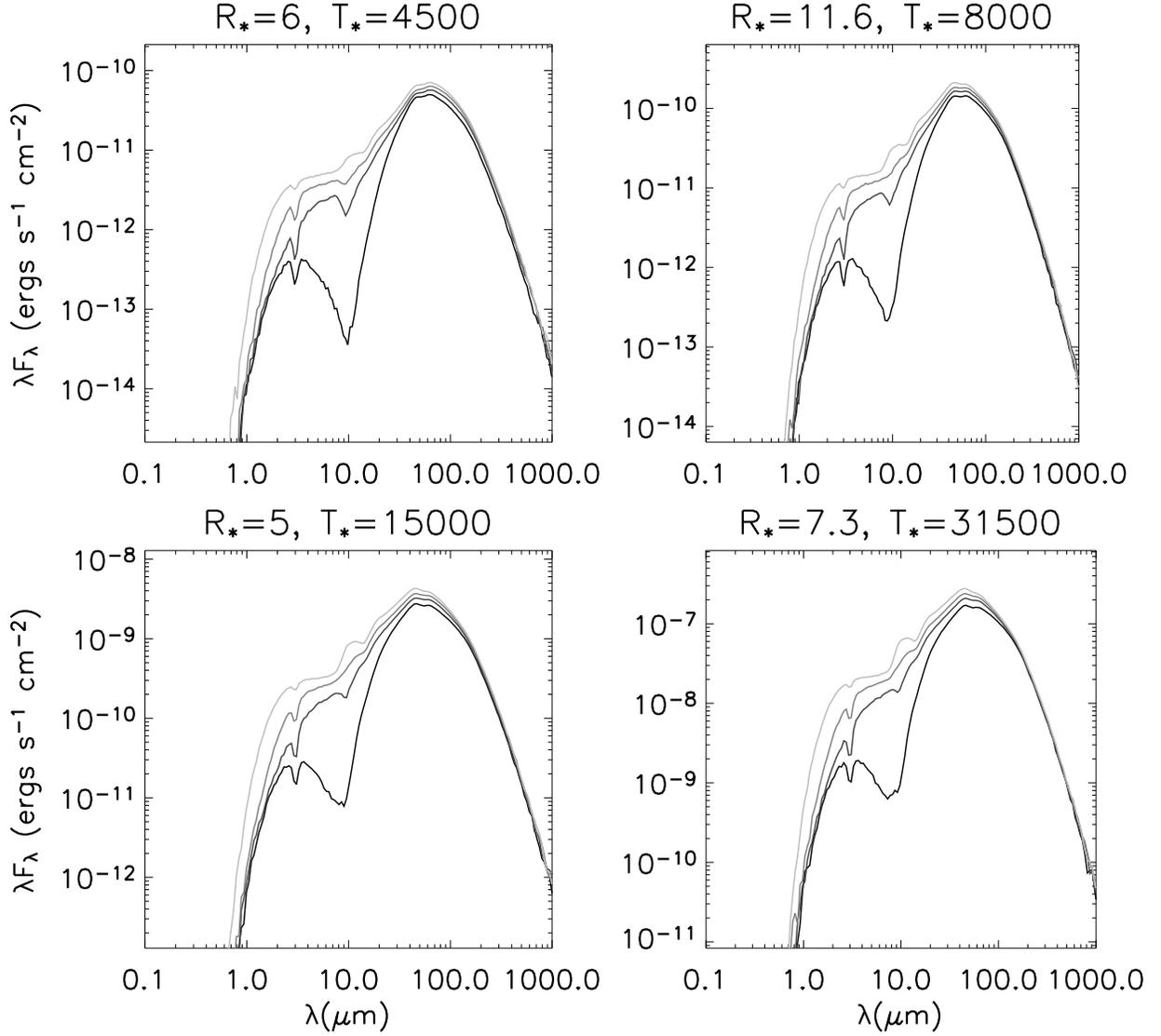}
\caption{\label{homoga} SEDs of the four homologous models.
Four inclinations are plotted: $\cos i = 0.05$ (edge-on), 0.35, 0.65, 0.95 (pole-on).
Inclination variations are shown with different shades of grey from
light grey (pole-on) to black (edge-on).  Fluxes are scaled to a
distance of 2 kpc.  (a) The thermal spectrum only.  (b) total
spectrum, including stellar direct+scattered flux and thermal.  The
input stellar spectrum is shown as a dotted black line.  }
\end{figure}

Figure 2 shows SEDs for the homologous model.  Figure \ref{homoga} shows only
the thermal spectrum from each model, and does not include the stellar
direct or scattered radiation.  From this we can see that the shapes of
the spectra are similar for all models.  The temperature structure and therefore thermal emission
is determined only by the luminosity of the incident radiation (which
sets the dust destruction radius) and where the radiation is absorbed
(determined by geometry and density). The slight differences between
the models are due to the fact that the models are not perfectly
homologous for two main reasons: 1) The geometries are slightly
different between the models because the ratio of the stellar size to
dust destruction radius varies: In the high-$T_\star$ models, the star is
effectively a point source, and in the low-$T_\star$ models it is larger
than the inner disk wall height; and 2) Because the dust scattering
albedo is non-zero and varies with wavelength, the total absorbed
luminosity varies slightly between the models (i.e., the dust scattering albedo at
the wavelengths where most of the stellar flux is emitted is lower in the
low-$T_\star$ models than the high-$T_\star$ models so a slightly higher
fraction of flux is absorbed in the low $T_\star$ models).  However, the
differences are slight and they do not detract from the main point,
that the thermal spectra are similar between the models.

\begin{figure}
\figurenum{2b}
\epsscale{1.0}
\plotone{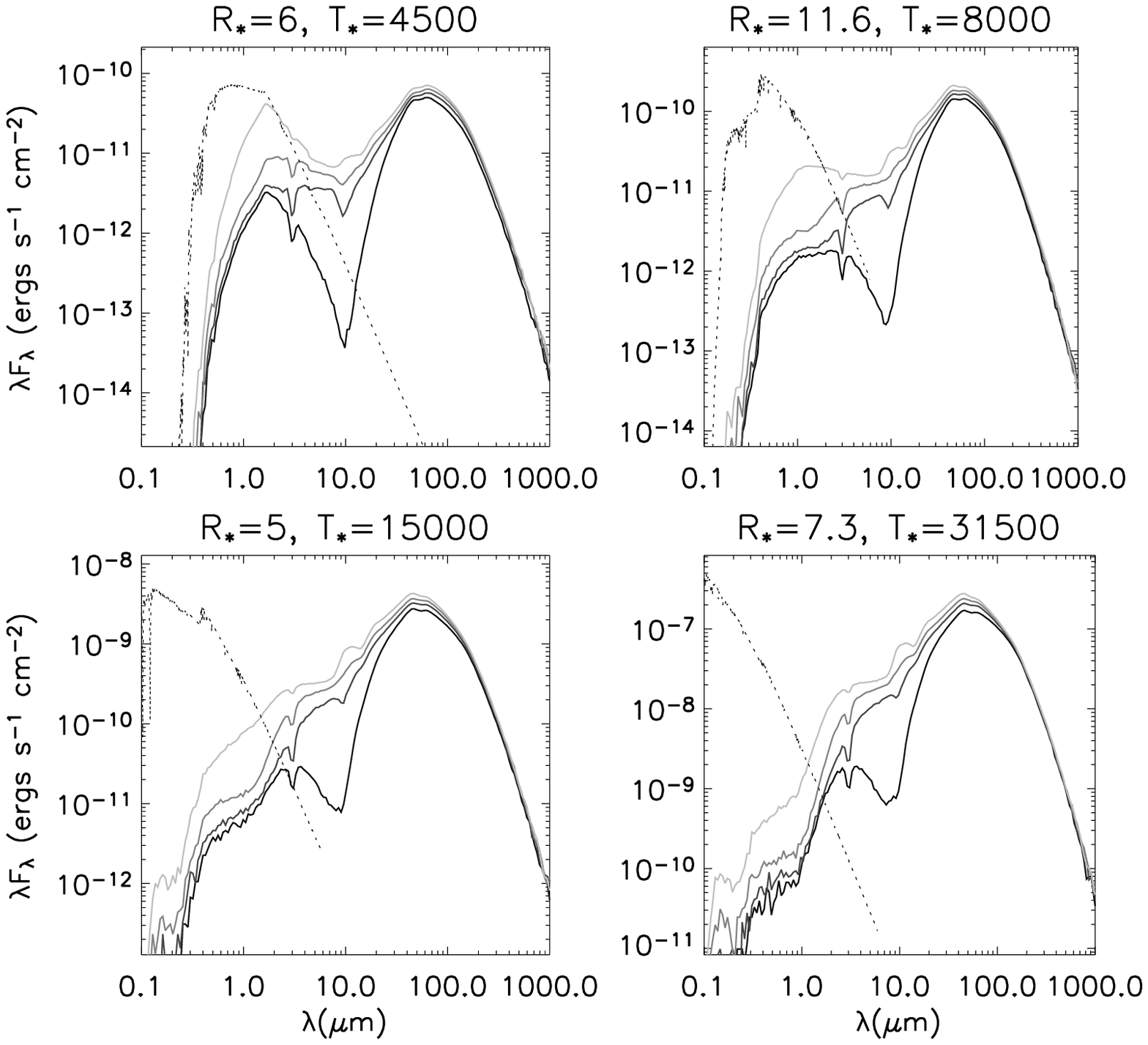}
\caption{\label{homogb}}
\end{figure}

Figure \ref{homogb} shows the total spectrum from each model.  Here we see that
the stellar contribution (scattered and direct) is greater in the
low-$T_\star$ models.  Thus, compared to the high-$T_\star$ models, the
spectrum is relatively blue.  We emphasize that the high-temperature
models are redder in the 1-10 $\mu$m region, not because the envelope mass is
larger but because the spectrum in this region is a more pure thermal
spectrum.  Note also that one reason the thermal spectra are similar
for all the models is that the dust sublimation temperature $T_{sub}$ sets the
cutoff for the maximum temperature of the thermal radiation.  Thus the
peak of the thermal emission occurs at the same wavelength for all the
models.  If there were no cutoff with $T_{sub}$, then the high $T_\star$
sources would have a more continuous blend between stellar and thermal
spectra.

\subsection{Class I Models guided by observations}

The high-$T_\star$ models in Figures 2 are not realistic,
given the homologous scaling of the disk parameters from the low-$T_\star$ disk. Here we
show more realistic geometries gleaned from previous observations and
modeling of high-mass sources
\citep[e.g.][]{alvarez04,beltran,beuther,sandell04,sandell00,shepherd}.  We keep the
envelope optical depths similar between the models to help understand the comparisons
between the models better.
The model parameters are shown in Table 2.  
The main difference between these and the homologous model is
in the disk parameters.  Like the homologous models, the disk inner radii
are set to the dust sublimation radius, $R_{sub}$.
The outer radii are chosen based on observations cited above.
The disk masses are chosen to be 5\% of the stellar mass.
The disk scale heights are calculated at $R_{sub}$ from the analytic solution of
the hydrostatic equation assuming the disk temperature is vertically isothermal.
Since the disk temperature is known at $R_{sub}$, it is straightforward
to calculate the scale height $h$ at this location (equal to the sound speed
divided by the Keplerian velocity; Bjorkman 1997).  The gaussian scale
height at each radius is then $h=h_{sub} (r/R_{sub})^{1.25}$.
Disk accretion
is included, but the effect on the SED is minor since the disk
accretion luminosity is relatively small for all of the models (Table 2).
The outer envelope radii are left large since the hotter
stars will heat up surrounding ambient material out to several pc.
Note that the envelope masses of the high-$T_\star$ models are large
due to the large outer radii.  However, the relevant parameter for the
radiative transfer is optical depth, which is similar between the models.
Therefore the variation of the thermal emission is due to circumstellar 
geometry.  

\begin{figure}
\figurenum{3a}
\epsscale{1.0}
\plotone{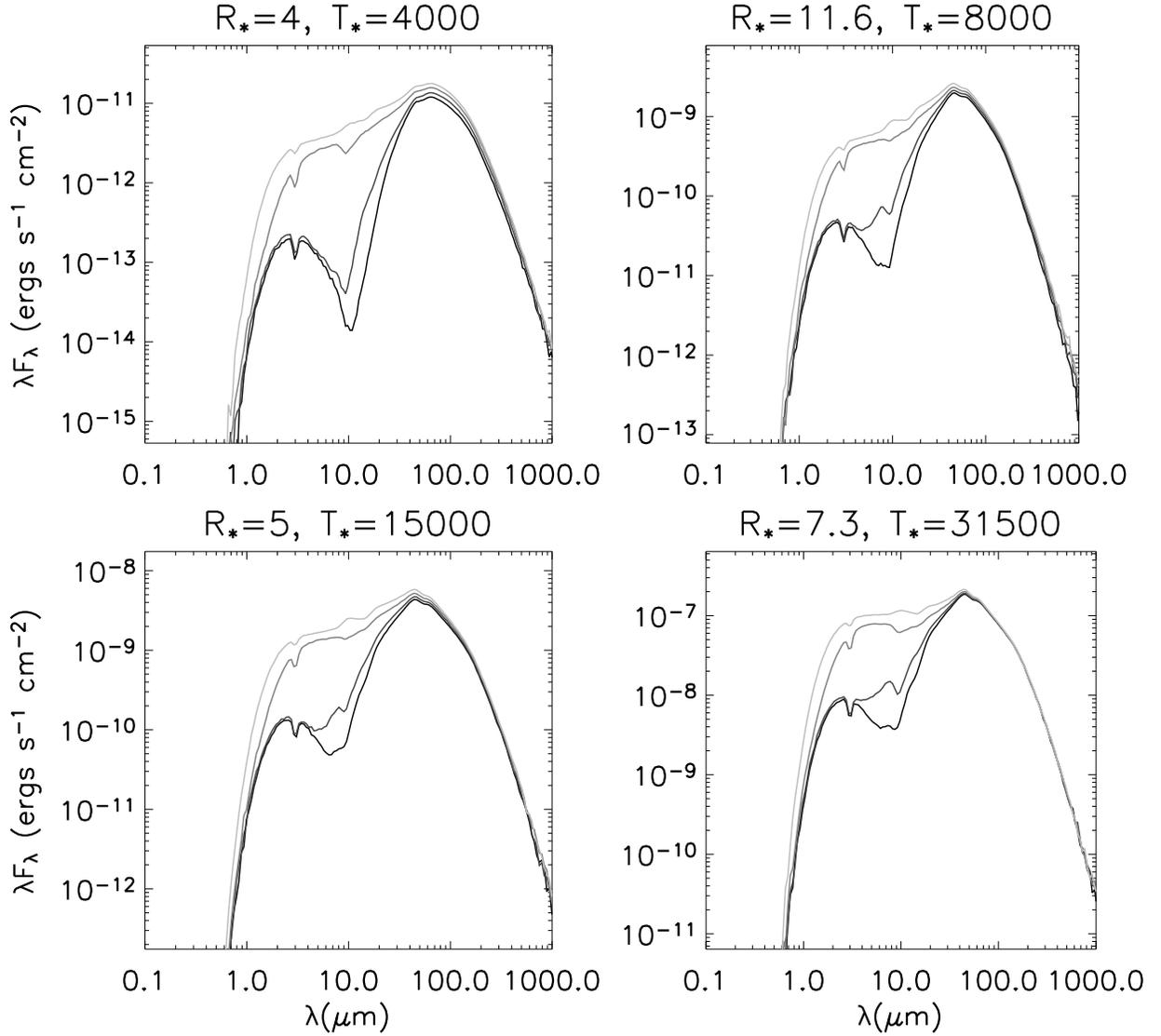}
\caption{\label{reala} SEDs of the four Class I models.  Inclinations
are the same as Fig. 1.   Fluxes are scaled to a distance of 2
kpc.  (a) The thermal spectrum only.  (b) total spectrum, including
stellar direct+scattered flux and thermal.  The input stellar spectrum
is shown as a dotted black line.  (c) the spectrum integrated in 3000
AU radius apertures (1.5\arcsec radius aperture at 2000 pc distance).
}
\end{figure}

\begin{figure}
\figurenum{3b}
\epsscale{1.0}
\plotone{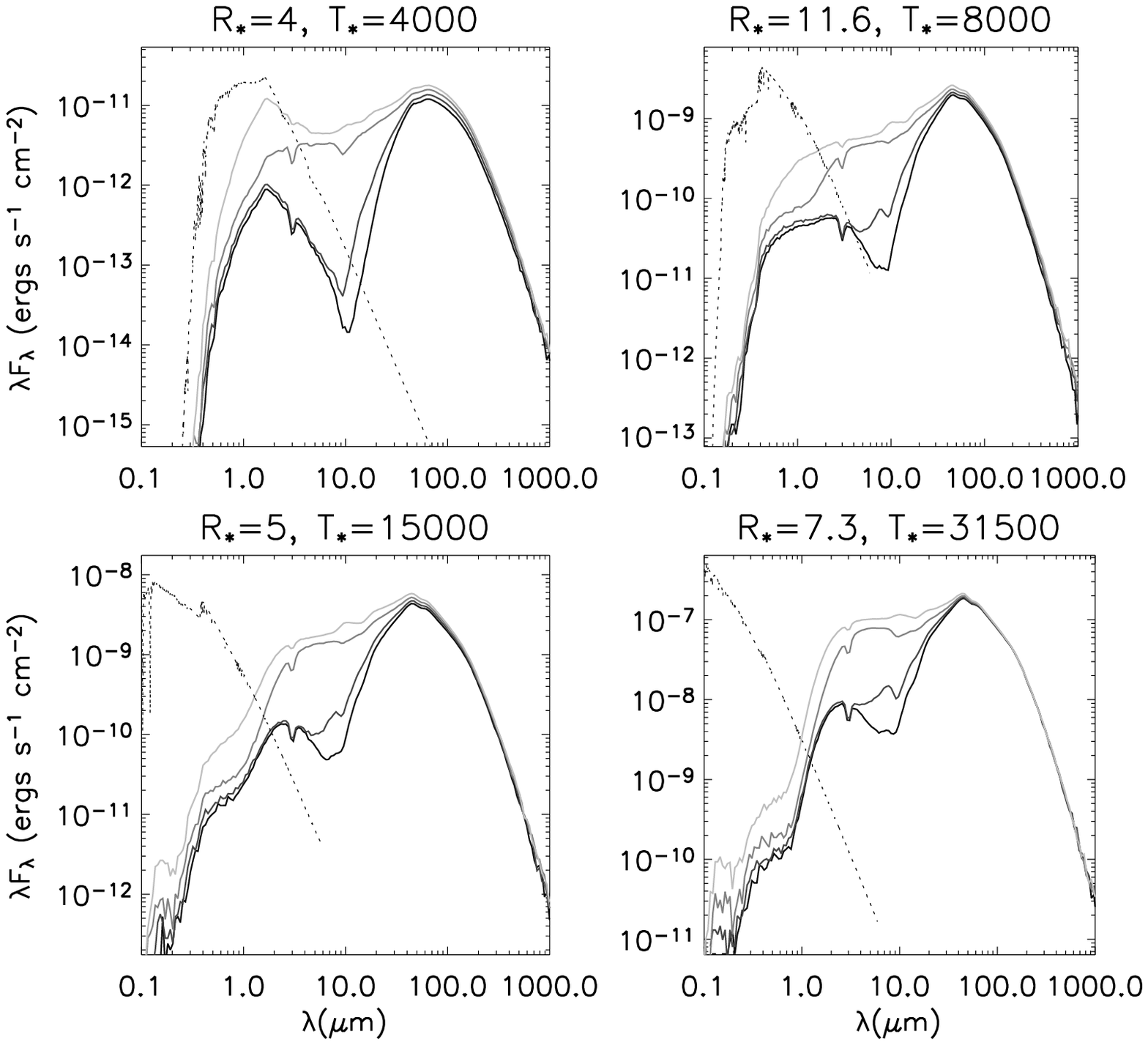}
\caption{\label{realb}
}
\end{figure}

\begin{figure}
\figurenum{3c}
\epsscale{1.0}
\plotone{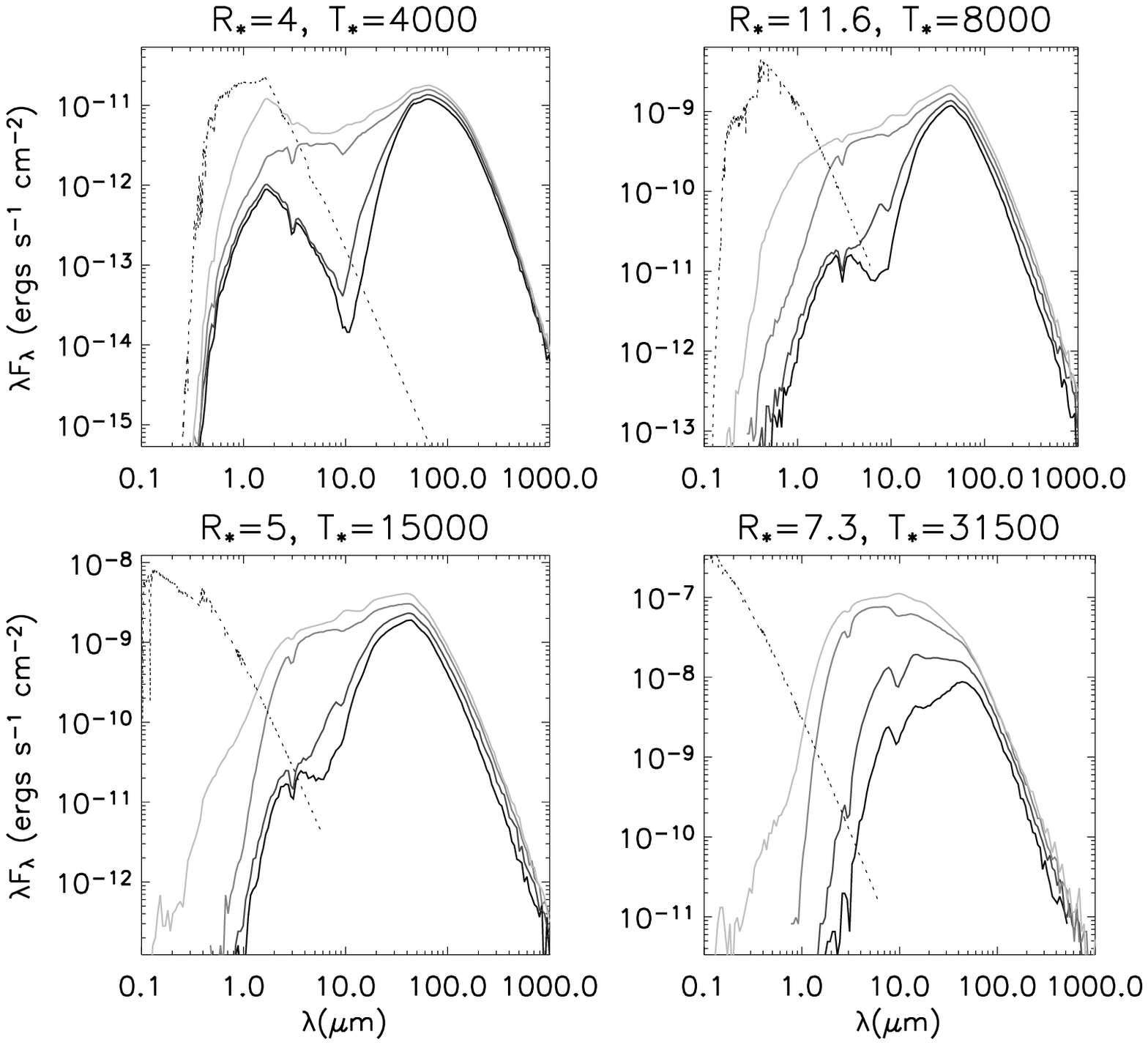}
\caption{\label{realc}
}
\end{figure}

Figure \ref{reala} shows
that hotter $T_\star$ models have relatively more $1-10 \mu$m thermal
emission than the cooler $T_\star$ models.  
This is due to the larger inner disk walls in the high-$T_\star$ models as a result of the larger
dust destruction radius ($h/r$ at $R_{sub}$ in Table 2).
This is similar to the ``puffed-up'' inner disk region invoked by Natta et al. (2001) to explain the 
SEDs of Ae/Be stars.
The wall intercepts and reprocesses radiation near the dust sublimation temperature
$T_{sub}$ with its peak radiation at about 2 $\mu$m.
This effect did not appear in the homologous model because 
$h/r$ at $R_{sub}$ was the same in all the models (Table 1).
The increased $1-10 \mu$m emission in the high-$T_\star$  models is
counteracted slightly when the stellar emission is included (Figure
\ref{realb}) but for the models with $T_\star > 8000$K, the colors are
bluer towards pole-on inclinations and higher $T_\star$ due to emission from the disk wall.
This figure shows that the disk structure and emission properties 
affect the 1-10$\mu$m spectrum in Class I sources.
Note that the SEDs in Figs. \ref{reala} and \ref{realb} include the flux from the entire envelope which
extends to nearly 2 pc in the case of the hot star model (Table 1).  
Figure \ref{realc} shows the results
integrated in a 3000 AU radius aperture (1.5\arcsec at a distance of 2
kpc), more typical of aperture photometry observations.  
In this case, the hottest $T_\star$ model has less short- and
long-wave flux in the smaller aperture giving a more rounded SED
shape.  It is rather striking that the SEDs of these four Class I
models, with nearly identical optical depths in the cavity and envelope, 
have such different shapes.

\subsection{Other Evolutionary States:  Class 0 and II}

\begin{figure}
\figurenum{4}
\epsscale{1.0}
\plotone{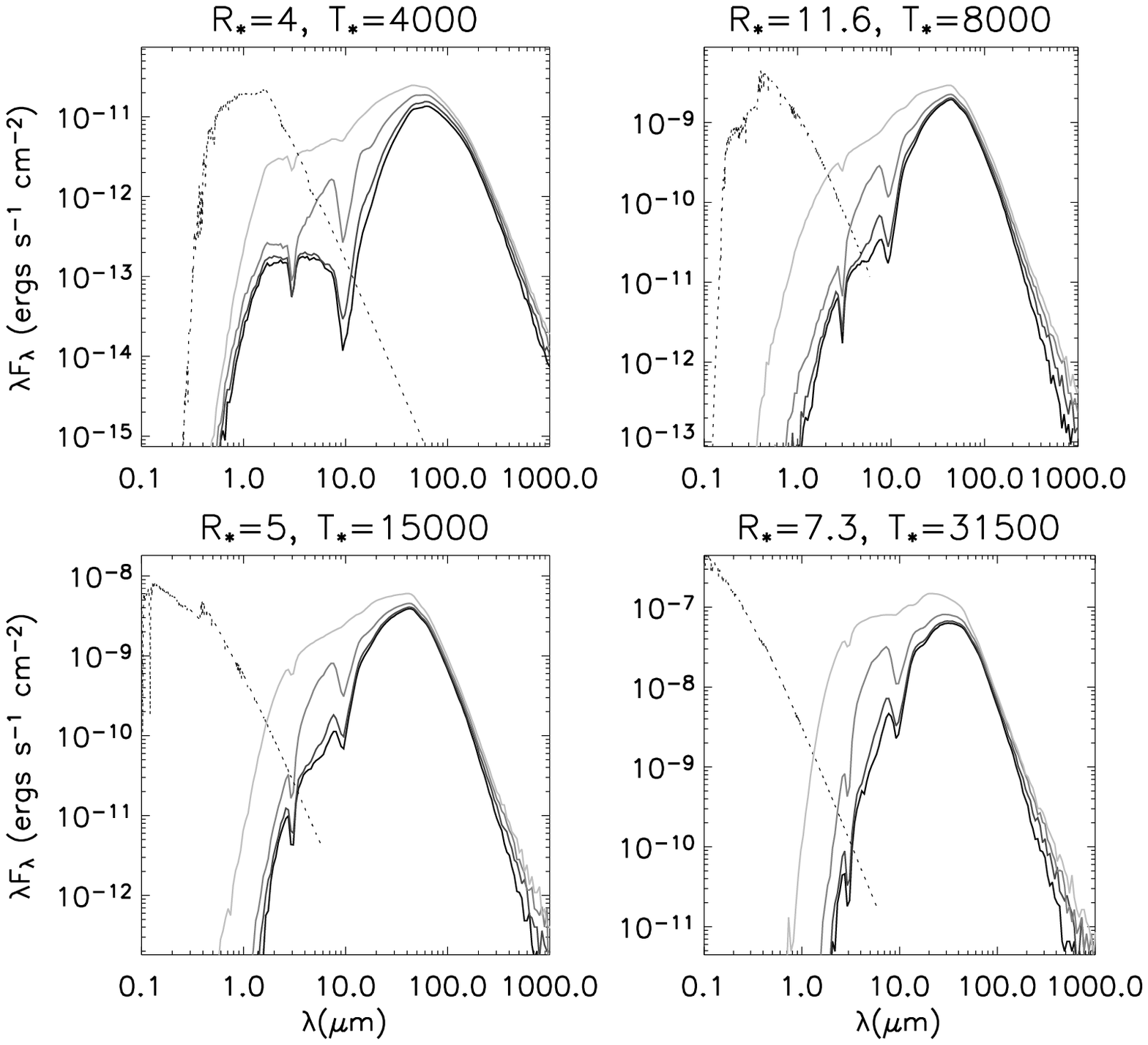}
\caption{\label{c0}  Class 0 SEDs (Table 3). The total spectrum (stellar direct+scattered
and thermal) is plotted for each stellar temperature.  Inclination variations are as
in Figure 1.  Fluxes are scaled to a distance of 2 kpc.  
The spectra are integrated over 3000 AU radius apertures from the central source. }
\end{figure}

To see how stellar temperature affects other evolutionary
states, we also show SEDs of Class 0 and II sources for
the four stellar temperatures.
We keep the stellar parameters the same for the Class 0 and II models
even though they would obviously evolve over this time period.
However, this allows us to isolate better the differences between the resulting
SEDs (e.g., the disk scales heights will be similar if the stellar properties
do not change).
Figure~\ref{c0} shows SEDs for Class 0 sources with the model parameters
in Table 3.  The envelope optical depths are four times higher than the Class I
model, and the polar optical depths are two times higher. 
The disk radii are smaller and disk masses higher (0.1 times the stellar mass),
giving larger midplane optical depths.
The SEDs are shown integrated in a 3000 AU aperture.
Except for the pole-on inclinations, these show a trend for the high-$T_\star$
models to have redder $1-10\mu$m colors.
As in the Class I models, the disk walls contribute more radiation at 2-10$\mu$m in the high-$T_\star$
models but this is only apparent towards pole-on inclinations ($i < 45$\arcdeg).

\begin{figure}
\figurenum{5}
\epsscale{1.0}
\plotone{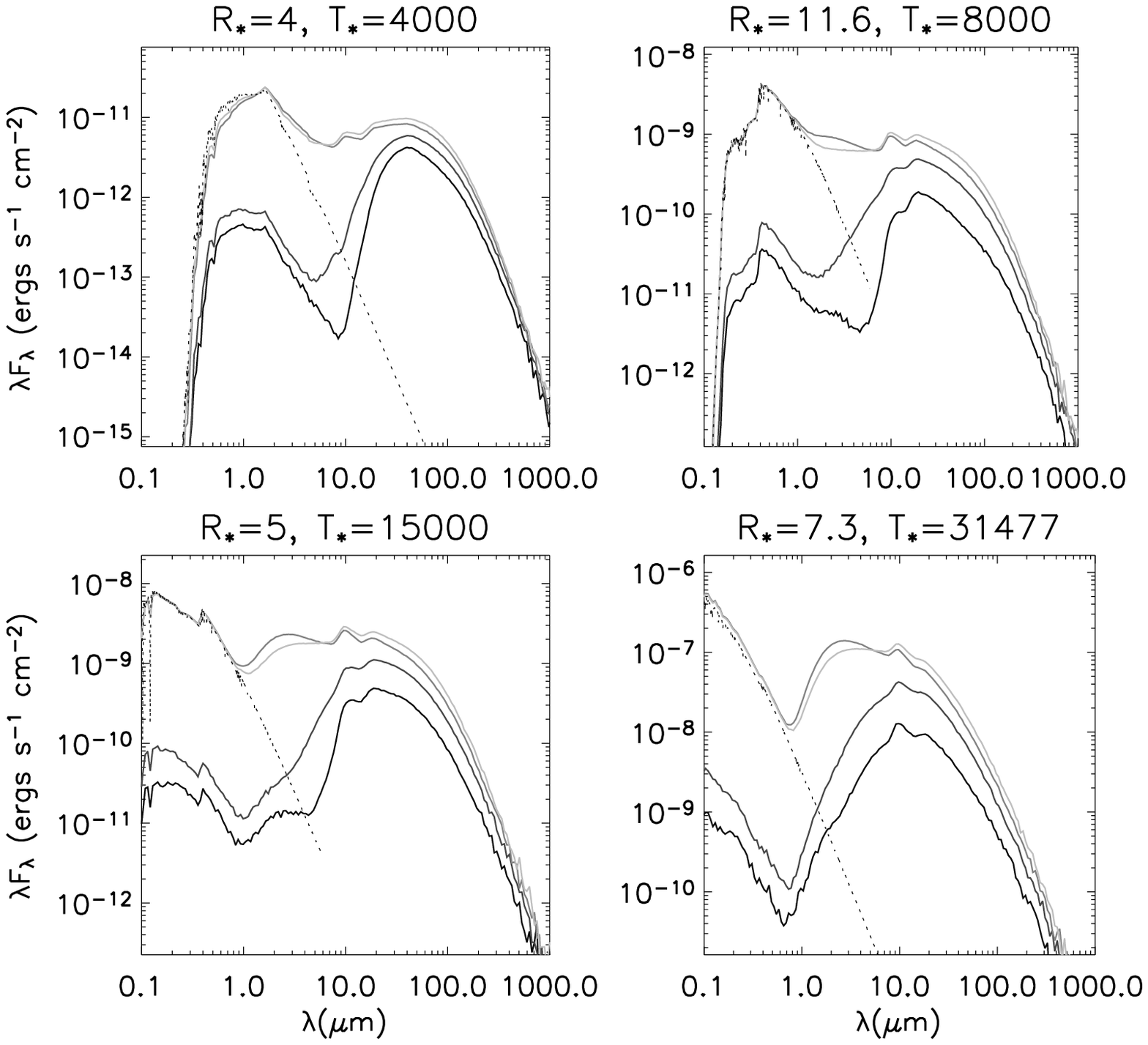}
\caption{\label{cii}  Class II SEDs (Table 4). The total spectrum (stellar direct+scattered
and thermal) is plotted for each stellar temperature.  Inclination variations are as
in Figure 1.  Fluxes are scaled to a distance of 2 kpc. 
}
\end{figure}

Class II SEDs are shown in Figure~\ref{cii} with the model parameters in Table 4.  
The disk parameters are similar to the Class I models except that the masses
are lower, since they are more evolved.   
At a wavelength range of 1-2$\mu$m, there is a tendency for the models to
become more red with increasing $T_\star$.
At 2 $\mu$m the disk emission kicks in and the larger walls from the high-$T_\star$
models show a blue spectrum towards pole-on inclinations from $2-10 \mu$m.  
However, the edge-on inclinations are red or flat in the models due to obscuration
by the flaring outer regions.
Thus when inclination is included, there is again a large spread in colors
at $1-10 \mu$m.
We note that the disk models do not include the effect of gas opacity inside
the dust destruction radius. 
This should not be a problem in the low mass disks due to low opacities
(Lada \& Adams 1992)  but it may be important in the high-mass
disks.  In addition, PAH emission is likely an important contributor in the high-$T_\star$
models.
This is beyond the scope of this paper and will be explored in the future. 

\subsection{Color-color and Color-magnitude diagrams}

\begin{figure}
\figurenum{6}
\epsscale{1.0}
\plotone{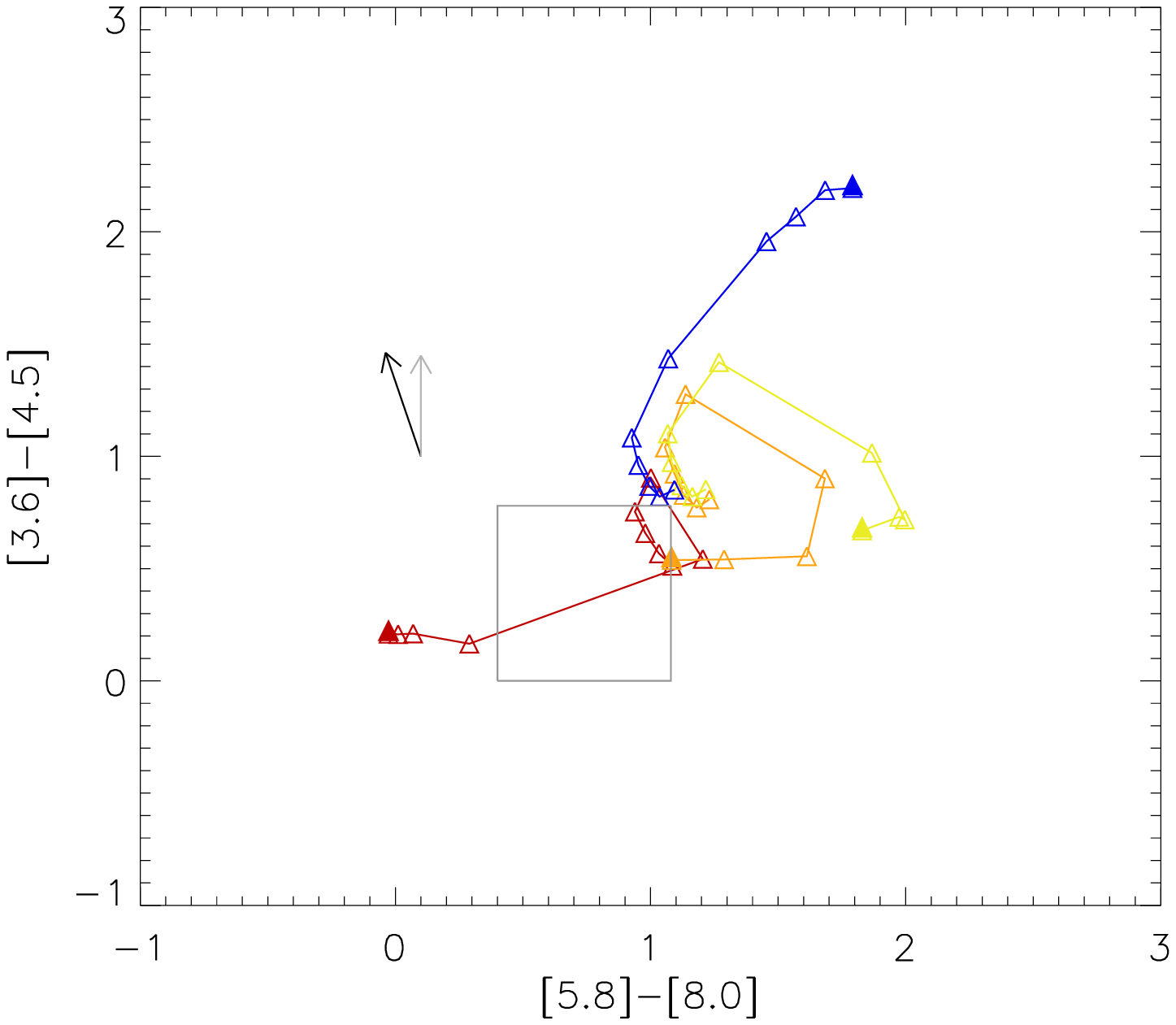}
\caption{\label{cc} 
Color-color plot ([5.8]-[8.0] vs. [3.6]-[4.5]) of the four Class I models
with fluxes integrated over a 3000 AU radius aperture.
The different colors are for the different stellar temperatures (red: T=4000 K;
orange: T=8000 K; yellow:  T=16000 K;  blue:  T=31500 K).
For each model, 10 inclinations ranging from edge-on to pole-on are plotted.  
The filled symbols show the edge-on colors.
The grey box is a region denoted by Allen et al. (2004) as the approximate
domain of Class II sources.
The black arrow denotes the reddening vector for the diffuse ISM (Cardelli, Clayton, \& Mathis 1989), and the grey arrow is derived from GLIMPSE data (Indebetouw et al. 2004).
The length of the arrows represent A$_V = 30$.
}
\end{figure}

\begin{figure}
\figurenum{7}
\epsscale{1.0}
\plotone{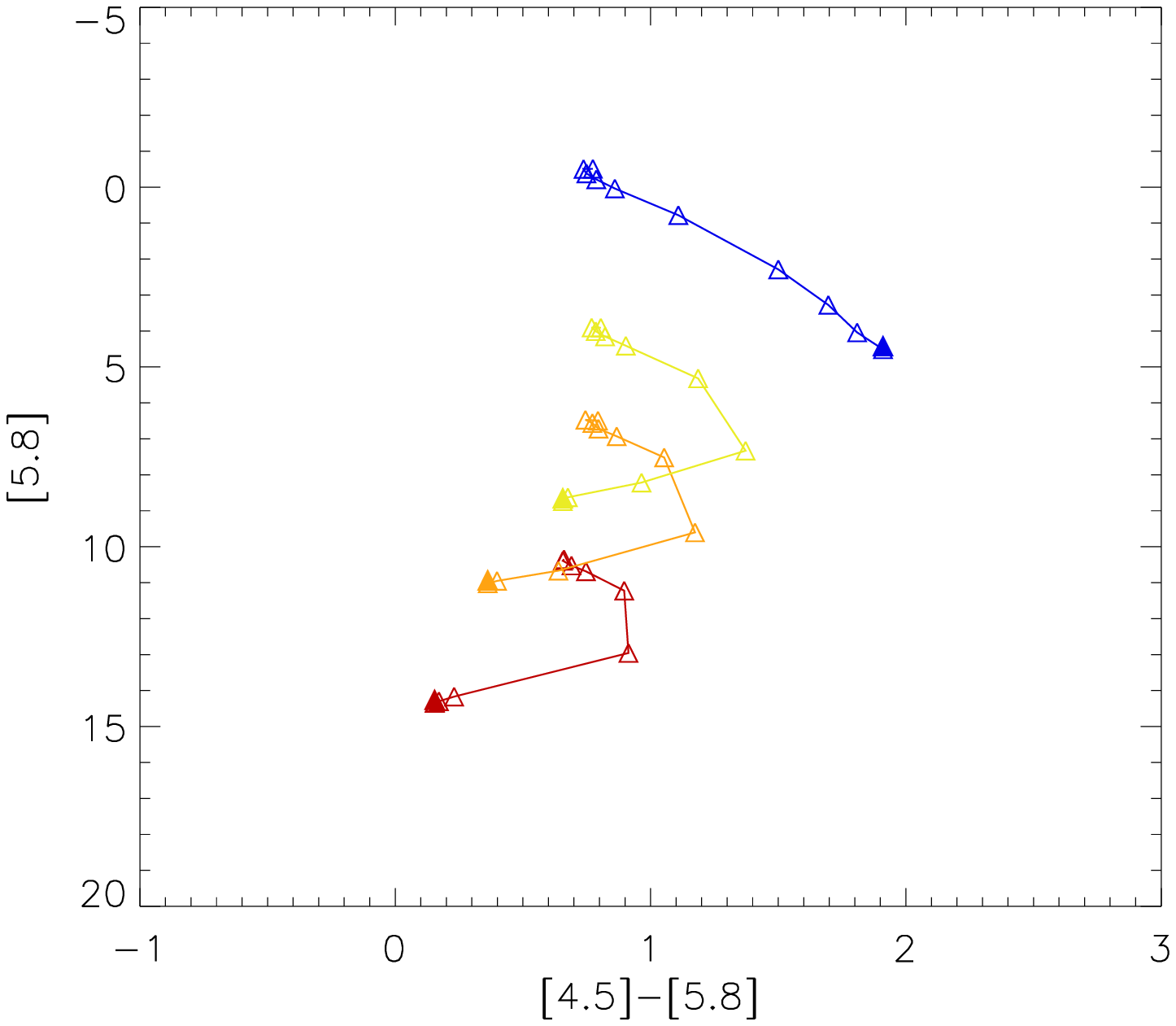}
\caption{\label{cm} Color-magnitude plot for the four Class I models
with fluxes integrated over a 3000 AU radius aperture.  The symbol colors
are as described in Figure~\ref{cc}.  The higher temperature sources
are more luminous (except for the edge-on sources in some cases).  Magnitudes are
scaled to a distance of 2 kpc.  }
\end{figure}

Figure \ref{cc} shows color-color plots in the IRAC bands [3.6]-[4.5]
vs. [5.8]-[8.0] for the Class I models (\S 2.3 and Table 2) integrated
in a 3000~AU radius aperture.  
There is a trend for the high-$T_\star$ models to be more red, though there
is overlap due to the broad spread in inclination within a model.
The color range of these Class I models (with similar optical
depths) spans the range of observations in the four star forming clusters
presented by \citet{allen04}!   
The grey box in Figure \ref{cc} shows the region denoted by Allen et al.
as the approximate domain of Class II sources.  
Some of edge-on Class I models are blueward of this domain (see Figure 3c).
Figure \ref{cm} shows a color-magnitude diagram for the Class I models.  This shows some
spread in both magnitude and color with each model but a general trend for the high-$T_\star$ models to be
brighter (obviously) and more red.  
The Class I models in Figure 6 can be compared to those presented by Allen et al.
(2004) in their Figure 1.
Their models show a trend for higher luminosity
sources to be more red in [5.8]-[8.0] and higher density envelopes (presumably 
younger sources) to be more red in [3.6]-[4.5].
It is not clear if they varied stellar temperature in their Class I models, but our
models show a similar trend for higher luminosity sources to have redder
[3.6]-[4.5] colors due to stellar temperature effects.  
The main difference however is that our models
in general give bluer colors for the same envelope parameters (density or
infall rate) due to our inclusion of partially evacuated bipolar cavities and
flared disks in the Class I models (Paper I, Figure 12).
Thus we would likely estimate a younger evolutionary state (higher density envelope)
for a given set of observational colors.

\begin{figure}
\figurenum{8}
\epsscale{1.0}
\plotone{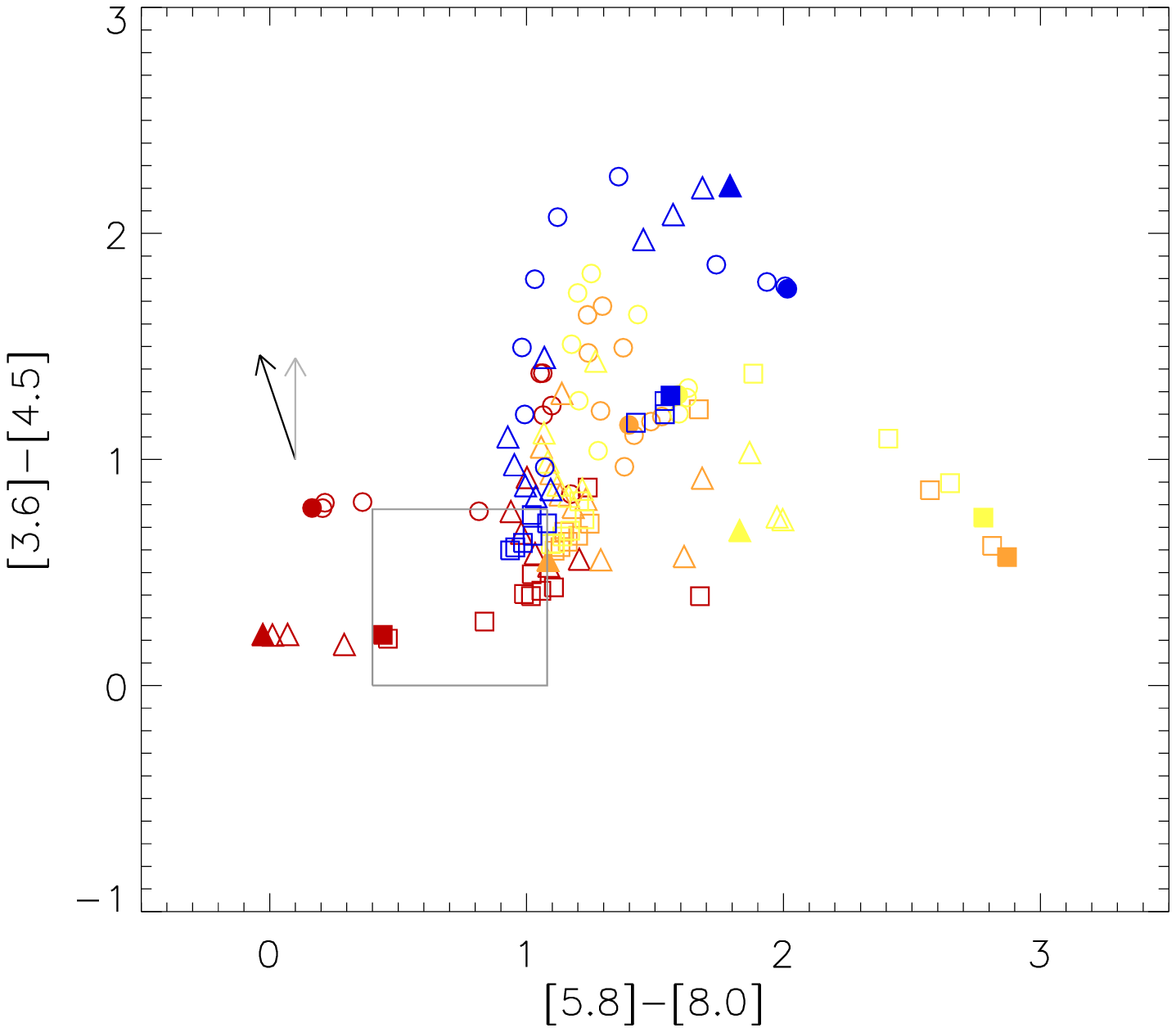}
\caption{\label{ccall} 
 Color-color plot including Class 0 and Class II
sources in addition to the Class I sources shown in Fig.~\ref{cc}. 
The different colors represent the different stellar
temperatures (red: T=4000 K; orange: T=8000 K; yellow: T=16000 K; blue:
T=31500 K).  The symbols represent the three evolutionary states
(circles: Class 0; triangles: Class I; boxes: Class II). 
For each model, 10 inclinations are plotted ranging from edge-on to
pole-on in equal intervals of $\cos i$.  The filled symbols show the edge-on sources.
The grey box is a region denoted by Allen et al. (2004) as the approximate
domain of Class II sources.
There is a
slight trend with higher temperature sources being more red, especially
in [3.6]-[4.5].  Notice
that five reddest sources in [5.8]-[8.0] are Class II sources.
The eight bluest source in [5.8]-[8.0] are Class 0 and I sources.
The arrows represent reddening vectors with A$_V = 30$ as described in Figure 6.
}
\end{figure}

\begin{figure}
\figurenum{9}
\epsscale{1.0}
\plotone{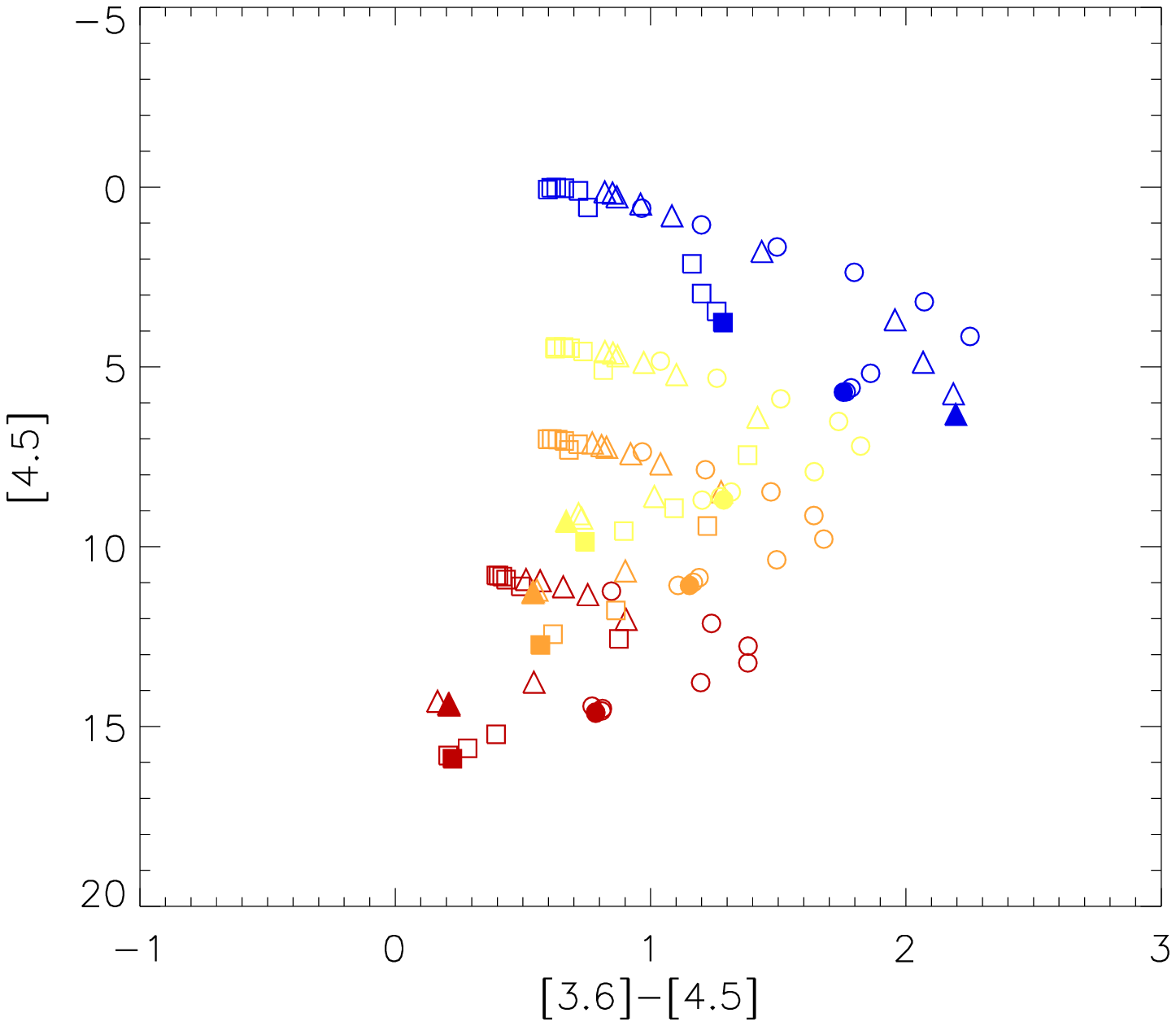}
\caption{\label{cmall}  Color-magnitude plot for all evolutionary states
and inclinations.  
The symbols are as
described in Fig.~\ref{ccall}.  
Magnitudes are scaled to a distance of 2 kpc.
For a given stellar temperature, the [3.6]-[4.5] color generally shows the expected
behavior with evolutionary sequence with the Class II sources at left
and Class 0 sources at right (with exceptions).
 }
\end{figure}

Finally, we show color-color and color-magnitude plots in Figures
\ref{ccall} and \ref{cmall} adding in Class 0 and Class II sources.
There are several interesting things to note in Fig.~\ref{ccall}:
the six reddest sources in [5.8]-[8.0] are Class II sources (with stellar
temperatures 8000 and 15000 K).   
The eight bluest sources in [5.8]-[8.0] are Class 0 and I sources (with
$T_\star=4000$ K).  
This is backwards from the common wisdom.
The [3.6]-[4.5] color behaves more as expected with Class 0-I sources
most red and Class I-II sources most blue.
There is somewhat of a dearth of sources in the ``Class II domain'' of Allen et al. (2004)
(the grey box).
This is due to the fact that we computed only one cool-$T_\star$ disk model,
whereas most T Tauri stars in a low-mass cloud are likely cool with a range of disk
masses.
Allen et al.'s disk models all used a stellar temperature of 4000 K and inclinations of
30\arcdeg and 60\arcdeg.
Both the hottest and coolest of our disk models fall at the right edge of their Class II domain
region.  The mid-temperature disks lie just to the right for most inclinations and
then to the far right for the partially obscured (near edge-on) sources.
Most of the Class 0 sources fall in the color range of 1-2 in both [5.8]-[8.0]
and [3.6]-[4.5].  
The Class I sources, on the other hand, span nearly the entire range of the
plot.  
They
suffer the most variation due to inclination, stellar temperature, inner disk
wall, and scattered
light effects.
And Class II and 0 sources can be found at unexpected locations in the color-color
plot as well, albeit with lower frequency.

Figure \ref{cmall} shows a more
systematic behavior in the color-magnitude plot.  The hotter $T_\star$
sources tend to be brighter, and within a given flux range, their
colors follow an evolutionary sequence with bluer colors being
younger.  Thus, the color-magnitude plot provides some guidance
in separating stellar temperature effects from evolutionary effects.

\section{Conclusions}

In this series of papers, we have shown that inclination effects, aperture
size, scattered light, and stellar temperature cause a broad spread in the colors of a source
at a single evolutionary state.   
There is systematic behavior, as shown in the color-magnitude diagrams (Fig. 9),
but the behavior is not as simple as using the slope
of the observed SED (or the IRAC [3.6]-[4.5] vs. [5.8]-[8.0] color-color
plot) to estimate evolutionary state (Fig. 8) for a given source.
There are trends in color space that could be applied in a statistical sense
to a cluster.   More modeling than the small grid presented here would be
useful to provide a statistical guide for interpreting mid-IR color-color plots.
Our codes are now publicly available and we are in the process of computing
a large grid of models which will also be publicly available.

We note that these problems for interpreting evolutionary state based on
SEDs are also reduced if long wavelength ($\lambda > 100 \mu$m) observations are
obtained.  These are much less sensitive to geometry (and inclination) and thus 1-D or
simple disk models do a reasonable job estimating circumstellar mass and hence
evolutionary state, assuming younger sources have more massive circumstellar
envelopes (e.g., Mueller et al. 2002).
Also, if the temperature of the stellar source can be estimated, the
problems of interpreting the mid-IR colors in terms of evolutionary
state become less severe.  
The stellar contribution can be estimated at wavelengths shortward of
2$\mu$m where disk emission does not contribute.
In our preliminary modeling of 
the {\it Spitzer} IRAC
data in the giant H II region RCW~49 (Whitney et al. 2004), we find
that we can estimate the stellar temperature by fitting multi-band
photometry across the $1-10 \mu$m range (2MASS JHK and the four IRAC bands).  
This improves our ability to
simultaneously estimate the evolutionary state and central source
temperature.
We will present these results in our next paper. 

\acknowledgments

This work was supported 
by the NASA LTSA Program (NAG5-8933, BAW);
by NASA's Spitzer Space Telescope Legacy
Science Program through Contract Number
1224653 (RI); by the National Science
Foundation (AST-0307686, JEB); and through a
UK PPARC Advanced Fellowship (KW).

\clearpage

\begin{deluxetable}{lllll}
\tablenum{1}
\tablewidth{0pt}
\tablecaption{Homologous Models}
\tablehead{
\colhead{Stellar Temperature:} &
\colhead{4000} & \colhead{8000} & \colhead{15000} &
\colhead{31500}
}
\startdata
Envelope infall rate ($/ 10^{-5} M_\sun/$yr) & 0.67 &  10.5 & 10  &  49 \\
Envelope mass ($M_\sun$) &  0.12 & 18 & 46 & 2150 \\
Stellar radius ($R_\sun$) & 4 & 11.6 & 5 & 7.3 \\
Stellar luminosity ($L_\sun$) & 3.67 & 494 & 1134 & 46900 \\
Stellar mass ($M_\sun$) & 0.5 & 6.0  & 6.2 & 20 \\
Envelope \& Disk inner radius ($R_\star$) & 6.7 & 29  & 106 & 500 \\
Envelope \& Disk inner radius (AU) & 0.125 & 1.55  & 2.47 & 16.9 \\
Envelope outer radius (AU) & 3000 & 36900 & 59000 & 404000 \\
Disk mass ($M_\sun$) & 0.01 & 1.28 & 3.06 & 138 \\
Disk outer radius (AU) & 300 & 3691 & 5900 & 40400 \\
$h/r$ at $R_\star$ & 0.011 & 0.0074 & 0.0053 & 0.0036 \\
$h/r$ at $R_{sub}$  & 0.017 & 0.017 & 0.017 & 0.017 \\
Cavity opening angle (\arcdeg) & 20 & 20 & 20 & 20 \\
Cavity density ($/ 10^{-20}$ gm cm$^{-3}$) & 37 & 3.0 & 1.9 &  0.28 \\
$\tau_V$ ($i=0$\arcdeg)   & 5 & 5 & 5 & 5 \\
$\tau_V$ ($i=60$\arcdeg) & 25 & 25 & 25 & 25 \\
$\tau_V$ ($i=90$\arcdeg, $/ 10^5$) & 1.93 & 1.93 & 1.93 & 1.93 \\

\enddata
\tablecomments{Inner Envelope/Disk radii are set to the dust sublimation radius, $R_{sub}$.
}
\end{deluxetable}

\clearpage

\begin{deluxetable}{lllll}
\tablenum{2}
\tablewidth{0pt}
\tablecaption{Class I Models}
\tablehead{
\colhead{Stellar Temperature:} &
\colhead{4000} & \colhead{8000} & \colhead{15000} &
\colhead{31500}
}
\startdata

Envelope infall rate ($/ 10^{-5}  M_\sun/$yr) & 0.67 & 3.8 &  4.8 & 22 \\
Envelope mass ($M_\sun$) &  0.12 &  8.5 & 31 & 1430 \\
Disk mass ($M_\sun$) & 0.02 & 0.30 & 0.31 & 1 \\
Disk accretion rate ($/ 10^{-8}  M_\sun/$yr) & 2.1 & 24 &  24 & 120 \\
Disk accretion luminosity ($/10^{-4} L_{acc}/L_\star $) & 17 & 1.4 & 0.38 & 0.022 \\
Disk outer radius (AU) & 300 & 400 & 500 & 500 \\
$h/r$ at $R_\star$ & 0.025 & 0.018 & 0.016 & 0.016 \\
$h/r$ at $R_{sub}$  & 0.041 & 0.041 & 0.051 & 0.074 \\
Cavity density ($/ 10^{-20}$ gm cm$^{-3}$) & 37 & 3.0 & 1.9 &  0.28 \\
$\tau_V$ ($i=90$\arcdeg, $/ 10^5$) & 1.63 & 1.77 & 0.80 & 0.28 \\

\enddata
\tablecomments{
Only parameters different from Table 1 are listed here.
}
\end{deluxetable}

\begin{deluxetable}{lllll}
\tablenum{3}
\tablewidth{0pt}
\tablecaption{Class 0 Models}
\tablehead{
\colhead{Stellar Temperature:} &
\colhead{4000} & \colhead{8000} & \colhead{15000} &
\colhead{31500}
}
\startdata

Envelope infall rate ($/ 10^{-5}  M_\sun/$yr) & 0.88 & 5.2 &  6.1 & 25 \\
Envelope mass ($M_\sun$) &  0.15 &  11.5 & 58 & 2700 \\
Disk mass ($M_\sun$) & 0.05 & 0.60 & 0.62 & 2 \\
Disk accretion rate ($/ 10^{-8}  M_\sun/$yr) & 32 & 250 &  250 & 1400 \\
Disk accretion luminosity ($/10^{-4} L_{acc}/L_\star $) & 251 & 14 & 4.0 & 0.25 \\
Disk outer radius (AU) & 50 & 80 & 100 & 100 \\
$h/r$ at $R_\star$ & 0.025 & 0.018 & 0.016 & 0.016 \\
$h/r$ at $R_{sub}$  & 0.040 & 0.041 & 0.051 & 0.074 \\
Cavity density ($/ 10^{-20}$ gm cm$^{-3}$) & 74 & 6.05 & 3.8 &  0.55 \\
Cavity opening angle (\arcdeg) & 10 & 10 & 10 & 10 \\
$\tau_V$ ($i=0$\arcdeg)   & 10 & 10 & 10 & 10 \\
$\tau_V$ ($i=60$\arcdeg) & 100 & 100 & 100 & 100 \\
$\tau_V$ ($i=90$\arcdeg, $/ 10^5$) & 25 & 18.6 & 7.9 & 2.87 \\

\enddata
\tablecomments{
Only parameters different from Table 1 are listed here. 
}
\end{deluxetable}

\begin{deluxetable}{lllll}
\tablenum{4}
\tablewidth{0pt}
\tablecaption{Class II Models}
\tablehead{
\colhead{Stellar Temperature:} &
\colhead{4000} & \colhead{8000} & \colhead{15000} &
\colhead{31500}
}
\startdata

Disk mass ($M_\sun$) & 0.01 & 0.15 & 0.15 & 1 \\
Disk accretion rate ($/ 10^{-8}  M_\sun/$yr) & 1.1 & 12 &  12 & 59 \\
Disk accretion luminosity ($/10^{-4} L_{acc}/L_\star $) & 8.3 & 0.68 & 0.19 & 0.001 \\
Disk outer radius (AU) & 300 & 400 & 500 & 500 \\
$h/r$ at $R_\star$ & 0.025 & 0.018 & 0.016 & 0.016 \\
$h/r$ at $R_{sub}$  & 0.041 & 0.041 & 0.051 & 0.074 \\
$\tau_V$ ($i=90$\arcdeg $/ 10^5$) & 0.82 & 0.89 & 0.39 & 0.14 \\

\enddata
\tablecomments{
Only parameters different from Table 1 are listed here.
The envelope infall rate is 0.
}
\end{deluxetable}

\clearpage

\clearpage

\clearpage

\clearpage

\clearpage

\clearpage

\clearpage

\clearpage

\end{document}